# Deep-Subwavelength Spatial Characterization of Angular Emission from Single-Crystal Au Plasmonic Ridge Nanoantennas

*Toon Coenen, Ernst Jan R. Vesseur, and Albert Polman*

Center for Nanophotonics, FOM Institute AMOLF

Science Park 104, 1098 XG, Amsterdam, The Netherlands

coenen@amolf.nl

**Abstract**

We use spatially and angle-resolved cathodoluminescence imaging spectroscopy to study, with deep subwavelength resolution, the radiation mechanism of single plasmonic ridge antennas with lengths ranging from 100 to 2000 nm. We measure the antenna's standing wave resonances up to the fifth order and measure the dispersion of the strongly confined guided plasmon mode. By directly detecting the emitted antenna radiation with a 2D CCD camera we are able to measure the angular emission patterns associated with each individual antenna resonance. We demonstrate that the shortest ridges can be modeled as a single point dipole emitter oriented either upward ($m$=0) or in-plane ($m$=1). The far-field emission pattern for longer antennas ($m$>2) is well described by two interfering in-plane point dipoles at the end facets giving rise to an angular fringe pattern, where the number of fringes increases as the antenna becomes longer. Taking advantage of the deep subwavelength excitation resolution of the cathodoluminescence technique, we are able to determine the antenna radiation pattern as function of excitation position. By including the phase of the radiating dipoles into our simple dipole model we completely reproduce this effect. This work demonstrates how angle-resolved cathodoluminescence spectroscopy can be used to fully determine the emission properties of subwavelength ridge antennas, which ultimately can be used for the design of more complex and efficient antenna structures.

## Introduction

Optical nanoantennas form an interface between localized nanoscale emitters and the far field. In transmission mode they have the ability to direct and enhance radiative emission. Vice versa, in reception mode they can strongly confine incident free-space light into a nanoscale volume.[1] Plasmonic nanoantennas in particular, have gained much interest recently, because of their small footprint and strong interaction with light. Potential applications include enhanced light absorption in solar cells,[2,3] nanoscale photodection,[4] surface-enhanced spectroscopy for biosensing,[5] and strong enhancement of non-linear effects.[6,7] Several antenna designs consisting of multiple components have been proposed in literature such as *e.g.* Yagi-Uda antennas composed of an array of coupled metal nanoparticles[8-12] or bowtie antennas consisting of two closely spaced triangular nanoparticles.[13] Importantly, the individual building blocks of these more complex designs (metallic nanoparticles in various shapes and sizes) can act as antennas as well.[14-16] Experimental characterization and physical understanding of these single components is of vital importance for designing more complex geometries, but is difficult to obtain due to their nanoscale dimensions. An example of an elemental single-component antenna structure is the plasmonic ridge antenna.[17] In such a structure, surface plasmon polaritons (SPPs) are confined to a ridge that is carved in the surface of a metal substrate. In this antenna geometry, the two end facets of the ridge act as mirrors defining a cavity in which SPPs are resonantly confined, similar to the resonances observed in metallic nanowires.[18-22] Due to their small size, these Fabry-Pérot cavities have a large free spectral range; the cavity quality factor $Q$ is determined by the SPP propagation loss as well as the reflection losses at the end facets. One advantage of the ridge antenna is that it is monolithically attached to the metal substrate making fabrication relatively simple. Moreover, nearly all of the antenna radiation is emitted into the upper hemisphere due to the highly reflective substrate, enabling efficient radiation collection.

Previously, we have studied the resonances of these ridge antennas using cathodoluminescence (CL) imaging spectroscopy.[17] Using a 30 keV electron beam as a point source to generate SPPs on the ridge we measured standing SPP waves at well defined wavelengths corresponding to the cavity resonances.

While these measurements clearly indicate the presence of antenna resonances, they do not provide insight into the radiation mechanism of these antennas. Here we use the CL technique that is now expanded by two unique features to fully identify this mechanism. First, by using a piezo-electrical mirror stage we are able to precisely position the mirror collecting the CL signal in such a way that its focus coincides with the ridge antenna, increasing the photon collection rate ~50 times compared to our previous work. With this enhanced signal we resolve the spatial intensity distribution of the standing wave detection to an extent that the full dispersion of the SPP ridge plasmons can be accurately determined. Second, by using the new angle-resolved detection capability of the CL system, we are able to directly determine the far-field radiation pattern of the ridge antenna for every antenna resonance. By scanning the electron beam across the antenna, the radiation profiles are determined as function of position with deep subwavelength precision. The high spatial resolution, sensitivity, and angular detection capabilities of our CL instrument enables characterization of the ridge antenna properties to an extent that is unrivalled by any other measurement technique.

## Results and Discussion

Ridge antennas were fabricated by inversely patterning a single-crystalline gold substrate using focused-ion-beam milling (FIB). The ridge lengths were in the range 100-2000 nm, varied in 100 nm increments. All ridges were 120 nm wide (width measured 65 nm above the substrate) and 130 nm high. Figure 1 shows a scanning electron micrograph of a 700 nm long ridge together with a schematic of the structure. The FIB-milled arena surrounding the antenna is made to have a gradually decreasing depth away from the antenna to prevent parasitic SPP reflections. To probe the cavity resonances we scan a 30 keV electron beam in 10 nm steps over the center of each ridge and collect a CL spectrum for each position (for details see methods section). Figures 2a-c show the CL emission intensity as function of position and wavelength for ridges with lengths of 100, 500 and 800 nm respectively. The color scale represents the fraction of photons emitted per incident electron per unit bandwidth. For the 100 nm long ridge we observe a strong peak in CL emission in the center of the ridge around 750 nm free space wavelength. The fact that this resonance is only excited efficiently in the center suggests that it has $m=0$ (azimuthal) symmetry.[12,23,24] For the 300 nm ridge we observe strong CL emission near the end facets of the ridge, corresponding to a $m=1$ 'dipolar' resonance for which $\lambda_{spp}/2$ fits into the resonator. Here $\lambda_{spp}$ is the SPP wavelength at a frequency corresponding to the free space wavelength in the figure. For the 800 nm ridge standing wave patterns are visible at $\lambda=600$ nm and 700 nm, with 4 and 3 antinodes respectively, corresponding to a $m=2$ 'quadrupolar' resonance and $m=3$ 'octupolar' resonance. Figure 2d shows the standing wave profile for these resonances, obtained by plotting the CL intensity of the data in Figures 2a-c along the ridge length at the resonance wavelength. The resonance orders $m=0,1,2$ and 3 are indicated in the figure. It is interesting to note that the CL technique enables probing of the even-order resonances. This is not possible with optical excitation under normal incidence, for which excitation of even order resonances is forbidden for symmetry reasons.[22,25] In contrast, the point-like electron beam excitation is not restricted to odd-order resonances.[26] The very high signal-to-noise ratio in the data of Figure 2d reflects the high collection efficiency for this antenna geometry and the sensitivity of our CL-system.

Next, 2D excitation maps are made by raster scanning the electron beam over the antenna in two dimensions. By using the resonance wavelengths identified from the linescans, we can plot the 2D spatial profiles of the resonances. Figures 2e-h show 2D CL maps for the 300 and 800 nm long ridges at the resonance wavelengths integrated over 10 nm bandwidth. Figure 2e shows the $m=1$ resonance at $\lambda=650$ nm for the 300 nm long ridge. This antenna possesses a $m=0$ resonance as well at $\lambda=950$ nm, for which the spatial profile is plotted in Figure 2f. The $m=0$ nature is clearly reflected in the uniform intensity observed across the entire antenna area. Figure 2g and h show the $m=3$ and $m=2$ resonance spatial emission patterns and 600 nm and 700 nm respectively. The spatial modulation of the emission intensity corresponding to the SPP standing wave pattern can clearly be distinguished.

By summing all CL spectra from a line scan we obtain an integrated CL spectrum showing all antenna resonances. Figure 3 shows such spectra for ridges with lengths from 100 to 1200 nm (we omit the longer ridges to reduce the number of curves but we did measure their spectra). The resonance orders $m$=0-5 are indicated in the figure. All resonances show a progressive red shift for increasing antenna length as expected. The visibility of the resonances decreases for longer antennas due to the increasing SPP propagation losses and the decreasing free spectral range. From the antinode spacing the plasmon wavelength can be deduced as this corresponds to $\lambda_{spp}/2$.[18] As the SPP experiences a phase shift upon reflection,[26,27] this analysis can only be performed on antennas with at least four antinodes ($m \geq 3$), such as in Figure 2g.

By using the standing wave profiles from multiple ridges for resonances with orders $m \geq 3$ we are able to reconstruct the complete dispersion curve for this plasmonic ridge antenna from 560-900 nm (1.4-2.2 eV). This result is shown in Figure 4. For reference we also show the light line in vacuum (black dashed line) and the dispersion curve for a SPP on a flat gold-vacuum interface (red curve), calculated using optical constants for single-crystal Au obtained with spectral ellipsometry. Figure 4 clearly shows that for the higher energies the measured dispersion strongly deviates from the planar SPP case and the light line; the largest effective modal index we find is $n_{spp}$=1.22 measured at $\lambda$=575 nm. This implies that the ridge waveguide mode cannot leak energy to SPPs on the substrate surrounding the ridge, nor can it couple directly to far-field radiation. Therefore we conclude that for these wavelengths most of the radiation is emitted from the antenna end facets as is illustrated in the inset of Figure 1. For larger wavelengths the dispersion in Figure 4 approaches the planar SPP dispersion and the light line indicating that at these wavelengths antenna radiation during propagation may contribute to the losses, possibly leading to a reduction in the $Q$ of the guided SPP mode. From Figure 2 it is clear that the end-facets are relatively bright compared to the central antinodes which can be attributed to a locally enhanced coupling to plasmons. This makes it difficult to extract the complex plasmon wave vector and reflection coefficients using a Fabry-Pérot model[22,26,28] since this enhanced coupling is not taken into account.

To obtain further insight into the confinement, loss and dispersion of the SPP modes we perform electromagnetic calculations using the two-dimensional boundary element method (BEM2D).[29,30] To obtain an exact representation of the geometry used in the experiment we made a cross section of one of the ridge antennas using FIB (Figure 5a) and we subsequently parameterized the SEM profile for use in the BEM calculations. In the calculation the waveguide is infinitely long, which means that effects from the end facets are not taken into account. We first calculate the induced near field on the ridge by a vertically oriented dipole located 10 nm above the ridge, mimicking the e-beam excitation,[12] for wavelengths of 600 nm (Figure 5b) and 900 nm (Figure 5c). In both cases the electric field intensity is clearly confined to the ridge. As expected the modal electric field is more confined at 600 nm than at 900 nm. Using BEM, we can also calculate the ridge mode dispersion by calculating the local density of states (LDOS) 10 nm above waveguide for different photon energies and mode wave vectors. Subsequently we fit a Lorentzian lineshape to the calculated LDOS *versus* wave vector for a given energy from which we derive the LDOS maxima to obtain a relation between wave vector and energy (see Figure 4, magenta curve). The calculated curve matches well with the experimentally determined dispersion data. From the width of the Lorentzian fits we can determine the modal $Q$, from which we can extract the propagation length ($L_{spp}$) of the waveguide mode.[17] We find for a free space wavelength of 600 nm that $Q = 16$ ($L_{spp}=1.3$ μm) and for 800 nm $Q = 130$ ($L_{spp} = 17$ μm). Using the same optical constants we calculate propagation lengths for planar SPPs of 5.4 μm and 49 μm respectively.

Since the waveguide propagation length is much shorter for higher energies than for lower energies we conclude that the waveguide loss in the detected wavelength range is dominated by ohmic damping rather than radiative damping. While these BEM calculations are made for infinite ridges, and thus only take into account propagation losses, the measurements are performed on finite-length ridges where reflection losses of the cavity also play an important role. As a result, the experimental resonance $Q$ will be lower than the modal $Q$. Indeed, the cavity $Q$ for all resonances shown in Figure 3 is estimated to be in the range of 2-10, significantly lower than the values calculated above. Moreover, the highest resonance $Q$'s are found for the shorter wavelengths, opposite to the trend based purely on propagation loss. This is explained by the fact that for shorter wavelengths the cavity end facet is a better mirror due to the higher modal index of the plasmon mode.

Using the complex plasmon wave vector calculated with BEM we can estimate the complex reflection coefficient by applying a simple Fresnel reflection model. Here the reflection coefficient $r$ is given by $(n_{spp}-1)/(n_{spp}+1)$ with $n_{spp}$ being the complex mode index $k_{spp}/k_0$.[27] If we do this for the 800 nm long ridge at $\lambda$=600 nm (m=3 resonance) we find that $r$ = 0.086 + 0.0153$i$. This shows that only a small fraction of the energy is reflected at the end facet and most of the energy is coupled to far-field radiation. From this coefficient we can also directly find the phase pickup $\Phi$ as $r = |r|e^{i\Phi}$ which for this antenna corresponds to 0.06$\pi$.[27] This is a relatively small phase shift compared to values reported in refs. 22 and 26. In those cases the metal antennas were ultrathin (25 and 20 nm respectively) leading to stronger plasmon confinement, higher ohmic losses and a larger mode mismatch with free space resulting in a much larger phase pickup which cannot be accurately determined with a simple Fresnel model.[27] An alternative method to estimate $\Phi$ is by measuring the spacing between the outer antinode at the end facet and the neighboring antinode. This distance is affected by the phase pickup, but as is clear from Figure 2g, the spacing is very close to the spacing between the central antinodes suggesting that there is only a minor phase pickup. In fact, if we calculate the phase pickup for the 800 nm ridge this way we find a value of 0.02$\pi$ which is close to the value predicted by the Fresnel model.

Using the angle-resolved detection capabilities in our CL instrument it is possible to determine the angular radiation profiles of the antenna, for every excitation position on the antenna, *i.e.* with deep subwavelength resolution. The CL emission is collected by the paraboloid mirror and imaged by a 2D CCD array from which we obtain the emission intensity as function of polar (zenithal) angle $\theta$ and azimuthal angle $\varphi$.[31] The measurements are made spectrally selective using 40-nm-band-pass color filters. We determine the angular emission distribution for antennas with different lengths at wavelengths corresponding to the antenna resonances. In this way, the radiation patterns for different resonance orders can be systematically compared.

We scan the electron beam in 25 nm steps over the center of the ridge in a line, and collect a radiation pattern at each position. The top row in Figure 6 shows the angular distribution of CL emission for *m*=0, 1, 2, 3, 4 and 5 obtained from 6 different antenna lengths (indicated in the figure). The depicted emission patterns were collected with the electron beam positioned on the central antinode for even-order resonances ($m$=0,2,4) and on one of the two central antinodes for the odd order resonances ($m$=1,3,5). The antennas were oriented normal to the optical axis of the mirror. For the shortest antenna (100 nm) the radiation pattern clearly has a toroidal 'doughnut' shape, consistent with the azimuthal symmetry associated with the $m$=0 resonance.[12,23,24] Since the ridge antenna is integrated with a conductive substrate this resonance is distinct from the vertically oriented dipole resonance in a metallic nanoparticle on a dielectric substrate as described in Ref. 9 where charge separation occurs in the direction normal to the surface. Instead, it can be attributed to a 'monopolar breathing type' resonance with a circularly symmetric charge distribution as shown schematically in the bottom row of Figure 6.[23] In terms of a Fabry-Pérot resonator it can be understood as being a "zero-length" cavity where the resonance condition is almost completely determined by the phase pickup upon reflection.[32] From the toroidally shaped emission pattern it is clear that this resonance has an effective dipole moment normal to the substrate. In contrast, the emission pattern for the $m$=1 resonance corresponds to that of a dipole oriented in the plane of the antenna as indicated by the charge distribution in the bottom row. The radiation profile from this dipole can be interpreted as an out-of-plane toroid with its symmetry axis oriented along the antenna's major axis, distorted by presence of the reflective gold surface, which leads to one strong radiation lobe pointing towards the surface normal. The next resonance, $m$=2, does not have a contribution normal to the surface but shows two off-normal emission lobes around $\theta$=30°. Subsequently, the $m$=3, 4, and 5 resonances show three, four, and five lobes respectively, where a central lobe pointing towards the normal is only observed for the odd-order resonances.

To model the antenna radiation patterns we distinguish between short and long antennas. The short 100 and 200 nm antennas were modeled by a single point dipole source that is oriented either vertically ($m$=0) or horizontally ($m$=1) along the antenna axis. We construct the far-field radiation pattern making use of the asymptotic far-field approximations discussed in Ref. 33 (section 10.6). We placed the point dipole 65 nm above the Au-surface, corresponding to the half of the antenna height. Optical constants for single-crystal Au from spectral ellipsometry were used as input for the calculations. The second row in Figure 6 shows the calculated angular radiation profiles. They show close resemblance to the measured profiles for the short antennas. To perform a quantitative comparison between experiment and theory we plot the polar distribution of the CL emission intensity by taking a cross cut along $\varphi=90°$ and $\varphi=270°$ (along the length of the antenna). The measured and calculated data, plotted in the third row of Figure 6, agree quite well.

For the longer antennas the point dipole approximation is not applicable, and the following model is developed. As described above, from the dispersion data it follows that, especially for shorter wavelengths, direct coupling of SPPs to the far field is weak and consequently the emission is mostly from the two end facets. This phenomenon has been observed experimentally by others as well.[34-36] Taminiau *et al.* have suggested that these types of travelling wave antennas can be approximated as two radiating dipoles oriented along the long axis of the antenna located at the end facets, with their phase and amplitude fully determined by the SPP waves.[28] We take a similar approach where the phase of the dipoles at the end facets is determined by the distance between the electron beam excitation position and the antenna ends and calculate the far-field radiation patterns of the two interfering dipoles for the relevant resonance wavelengths (see second and third row in Figure 6) by coherently adding the far-field of both dipoles.[33] The measured emission direction and the number of lobes are quite well reproduced by this relatively simple model. From the model we see that for $m=2,4$, with the end-facet dipoles 180° degrees out of phase, destructive interference occurs at $\theta=0°$. In contrast, for $m=3,5$ in-phase dipoles lead to constructive interference in that direction. For the $m=2$ resonance, good agreement between the calculated and measured angular patterns is observed. For the higher order resonances, the number of radiation lobes as well as the orientation is well described by our model, while discrepancies are found in the exact amplitudes. This is partly due to the fact that we used a 40 nm detection bandwidth in the experiment while the calculations were performed at the resonance peak wavelength. From the Rayleigh-Carson reciprocity theorem it follows that the even-order resonances cannot be excited with free-space light under normal incidence. In other words, if the antenna does not emit efficiently normal to the surface it is also not possible to efficiently excite it along that direction.

However, if excited under an angle, significant coupling to these even-order resonances is possible due to retardation effects, as is described in Ref. 22. In our experiment the electron beam acts as a vertically oriented point dipole source which does allow strong coupling. Thus, by using the CL technique we are able to efficiently probe the spectral and angular response of all the different resonance orders in the antennas, and directly see which resonances are difficult to access using optical far-field techniques.

To further test the far-field interference model, we investigate the radiation pattern for different excitation positions on the antenna. Our angle-resolved CL technique is a unique and ideal tool for this study, as it enables local excitation of the antenna mode with deep subwavelength spatial resolution for all wavelengths of interest. Figure 7 (top row) shows measured angular emission patterns for the $m=2$ resonance of the 500 nm long antenna at $\lambda=600$ nm collected at three different excitation positions as indicated in the electron micrographs in the bottom row. When excited in the center the dipoles are 180° out of phase giving rise to the symmetric radiation pattern found before in Figure 6. For excitation close to the end facets the pattern is no longer symmetric and the radiation emission is stronger away from the excitation point, along the antenna. Moving the electron beam across the antenna changes the relative phase of the dipoles since SPPs have to travel different distances to reach the two end facets resulting in asymmetric emission patterns.

For even-order resonances one might expect that the dipoles are always 180° out of phase, if excited in an antinode of the standing wave pattern. This is indeed the case for excitation at the central antinode, for which SPPs propagate an equal distance to both end facets and generate two oppositely oriented dipoles. If we would move the electron beam $\lambda_{spp}/2$ to another antinode, the dipoles should again be 180° out of phase, resulting in a symmetric angular emission pattern. However, for this quadrupolar resonance the last antinodes are offset in space due to the phase shift upon reflection, leading to a different relative phase between the dipoles and an asymmetry in the emission pattern. The second row in Figure 7 shows the calculated emission patterns using the dipole model and the third row again shows cross cuts through both the data and theory. The results are very similar to the measurement. The model predicts the same asymmetry as is observed in the measurement and also the magnitude of the lobes is predicted well.

## Conclusion

In conclusion, we have resolved the resonances and radiation patterns for gold ridge surface plasmon polariton nanoantennas with deep subwavelength resolution. From the measured antinode spacing in the spatial maps of the antenna resonances we determined the dispersion relation for the guided mode along the antenna; these results are in good agreement with boundary element method calculations. The resonance quality factor is in the range of 2-10 and is determined by ohmic damping within the metal and radiative losses at the end facets. The angular emission patterns for all resonance orders m=0-5 were resolved using angle-resolved cathodoluminescence spectroscopy. The emission pattern for antennas shorter than 200 nm can be modeled as a single point dipole source oriented either in the normal direction (100 nm antenna) or in the planar direction (200 nm). Longer ridges act as travelling wave antennas with a radiation profile that is well described by two radiating dipoles placed at the end facets, with their relative phase determined by the antenna length and the wavelength. Using the unique spatially resolved capabilities of the CL technique we were able to determine the antenna radiation pattern for different excitation positions. We find that changing the excitation position on the antenna changes the phase of the emitting end-facet dipoles and thus influences the radiation pattern. This study shows how angle-resolved cathodoluminescence spectroscopy can be used to probe nanoantenna properties in great detail, and with subwavelength precision. The fundamental insights obtained for this particular monolithic nanoantenna design can be used to design more complex coupled directional antenna structures with a tailored radiation profile.

**Methods**

**Antenna fabrication:** The ridges were fabricated with FIB using a FEI Helios dual beam system in a Czochralski grown single-crystalline gold pellet which was mechanically polished to achieve sub 10 nm RMS roughness. To fabricate a sharply defined gold ridge we used the lowest ion beam current of 1.5 pA and a short dwell time of 100 ns. The gradual change in depth of surrounding arena was achieved by slowly decreasing ion dose away from the antenna in a linear fashion. The cross section of the ridge was made by performing a cleaning cross section which prevents tapering. To ensure shape retention during milling a layer of platinum was deposited using electron beam induced deposition (EBID).

**Experimental setup:** The CL experiment was performed in a FEI XL-30 SFEG scanning electron micrope with an aluminum paraboloid mirror inside. The mirror alignment is performed using a specially designed piezoelectric positioning system.[12] For spectral imaging the CL was focused onto a fiber going to a spectrometer with liquid-nitrogen cooled CCD array. For the angular measurements the CL beam is projected onto a 1024x1024 imaging array. Emission patterns were obtained by mapping the 2D CCD image of the CL beam onto emission angle $\theta$ (zenithal angle) and $\varphi$ (azimuthal angle) and correcting for the collected solid angle per CCD pixel.[31]

**Cathodoluminescence experiments:** For the line scans a dwell time of 1s was used and for the spectral images 0.5s was used. For both a current of 1 nA was used. We corrected the data for system response by measuring the transition radiation spectrum for gold and comparing that to theory.[37,38] For the angle-resolved measurements we used 15s dwell time and a current of 10 nA to increase signal-to-noise ratio. Spectral sensitivity was achieved by filtering the CL beam with 40 nm band pass color filters. To purely look at the resonance emission from the antennas we subtract the transition radiation background. For both the spectroscopy and angular measurements we use a spatial drift correction algorithm to correct for the effects of beam and sample drift.

**Acknowledgements**

We would like to acknowledge Femius Koenderink for useful discussions. This work is part of the research program of the "Stichting voor Fundamenteel Onderzoek der Materie (FOM)", which is financially supported by the "Nederlandse Organisatie voor Wetenschappelijk Onderzoek (NWO)". This work was funded by an ERC Advanced Grant and is also part of NanoNextNL, a nanotechnology program funded by the Dutch ministry of economic affairs.

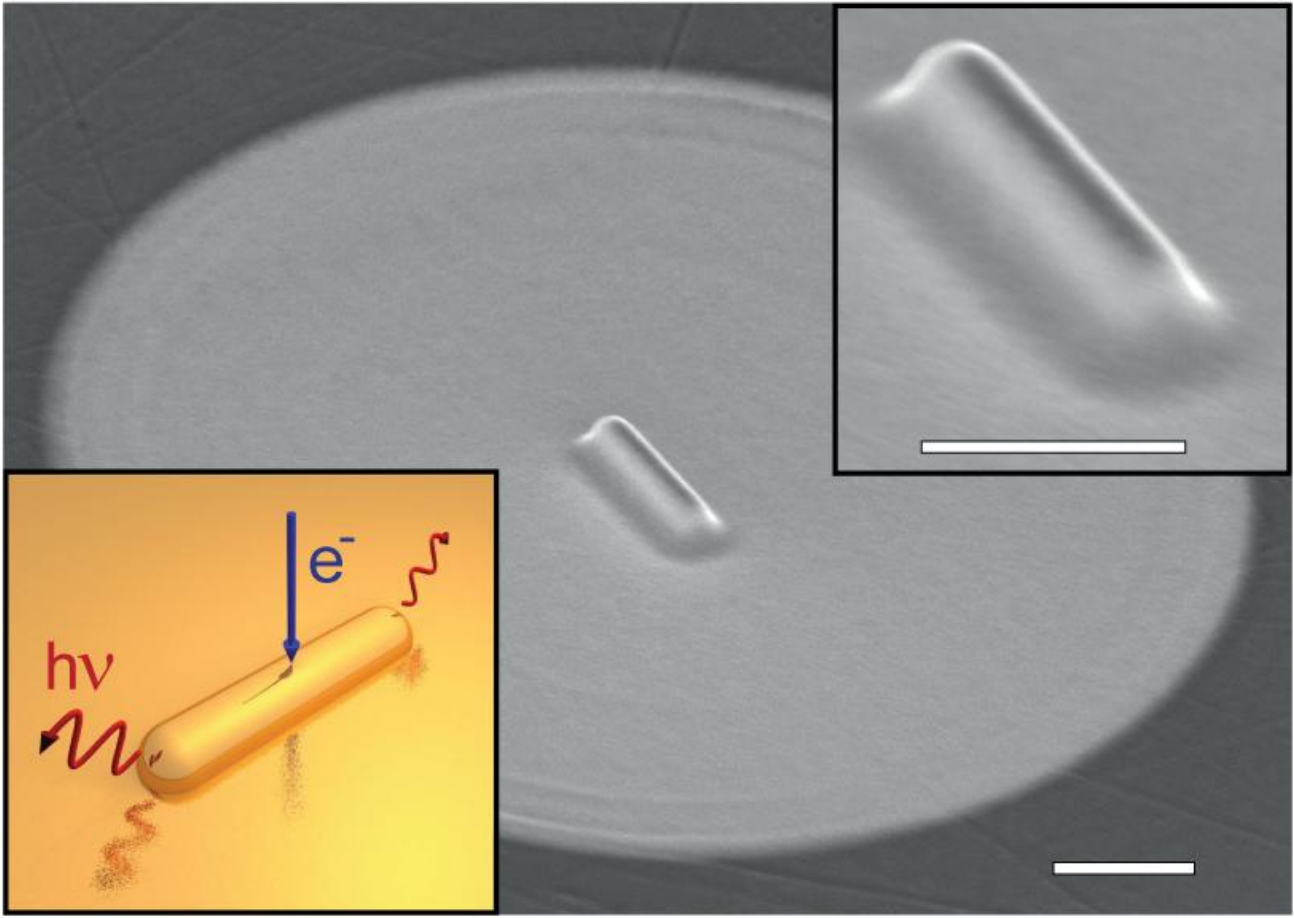

**Figure 1.** Scanning electron micrograph of a 700 nm long, 130 nm high and 120 nm wide (measured at 65 nm height) antenna. Top right inset: Close-up of the structure. Bottom left inset: graphic showing the e-beam excitation of the ridge with subsequent photon emission from the end facets. Scale bars are 500 nm.

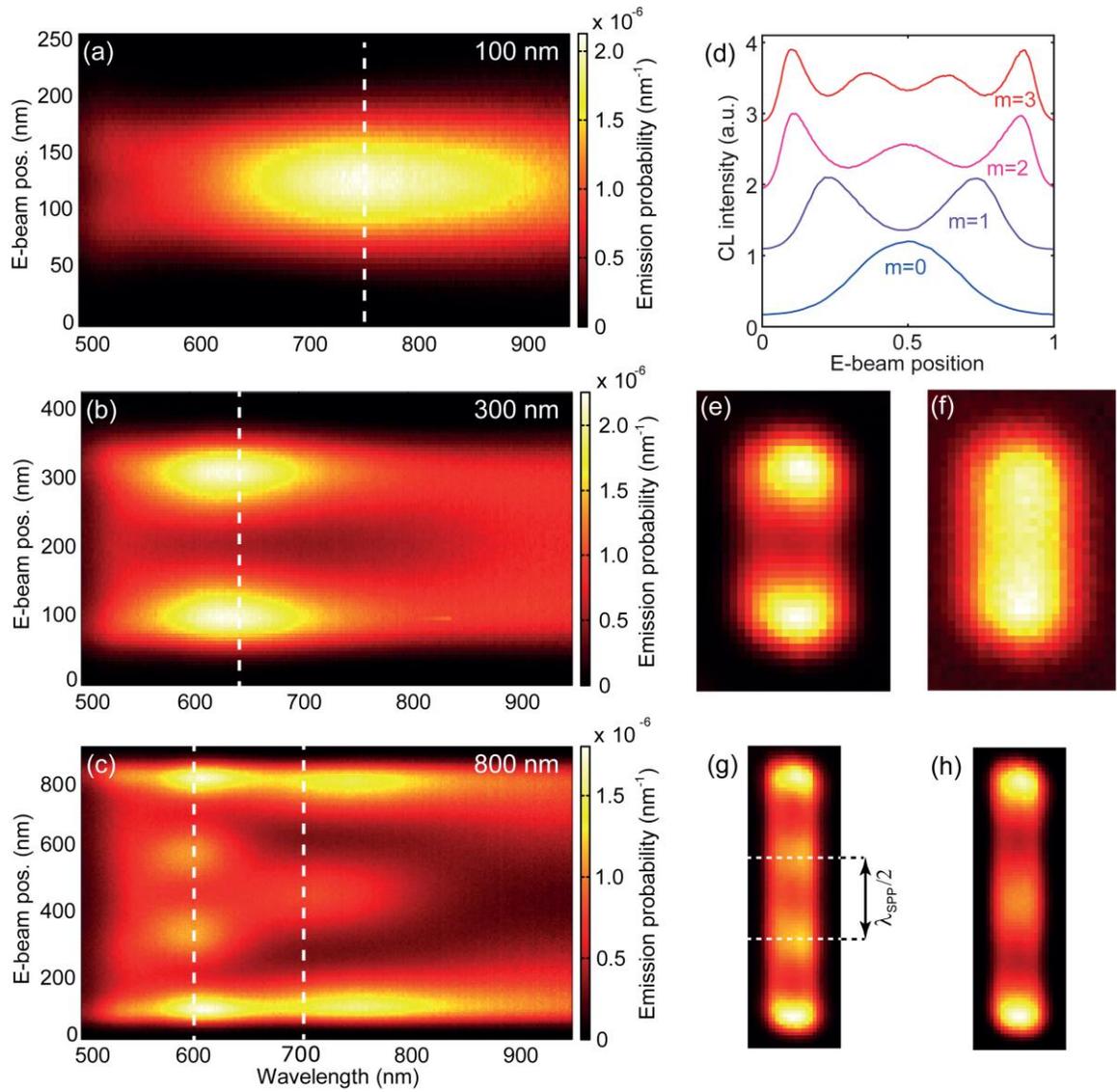

**Figure 2.** Experimental data: Collected CL emission probability per incident electron as function of position and wavelength, obtained by scanning the electron beam in 10 nm steps across a (a) 100 nm, (b) 300 nm and (c) 800 nm long Au ridge antenna. (d) Cross cuts through the experimental data along the dashed lines indicated in Figures 2a-c, showing the spatial resonance profiles for $m$=0-3. The CL-intensity has been normalized to 1 and the curves have been offset vertically for clarity. The horizontal scale has been normalized to the antenna length for each case. No data smoothing was used to obtain these curves. Spatially-resolved 2D excitation maps of the 300 nm long antenna, showing resonant CL intensity as function of e-beam position for (e) $\lambda$=650 nm ($m$=1) and (f) $\lambda$=950 nm ($m$=0) integrated over a 10 nm bandwidth. (g) Excitation maps for the 800 nm long antenna for $\lambda$=600 nm ($m$=3) and (h) $\lambda$=700 nm ($m$=2). For the $m$=3 resonance the spacing between the two central antinodes corresponds to $\lambda_{spp}/2$.

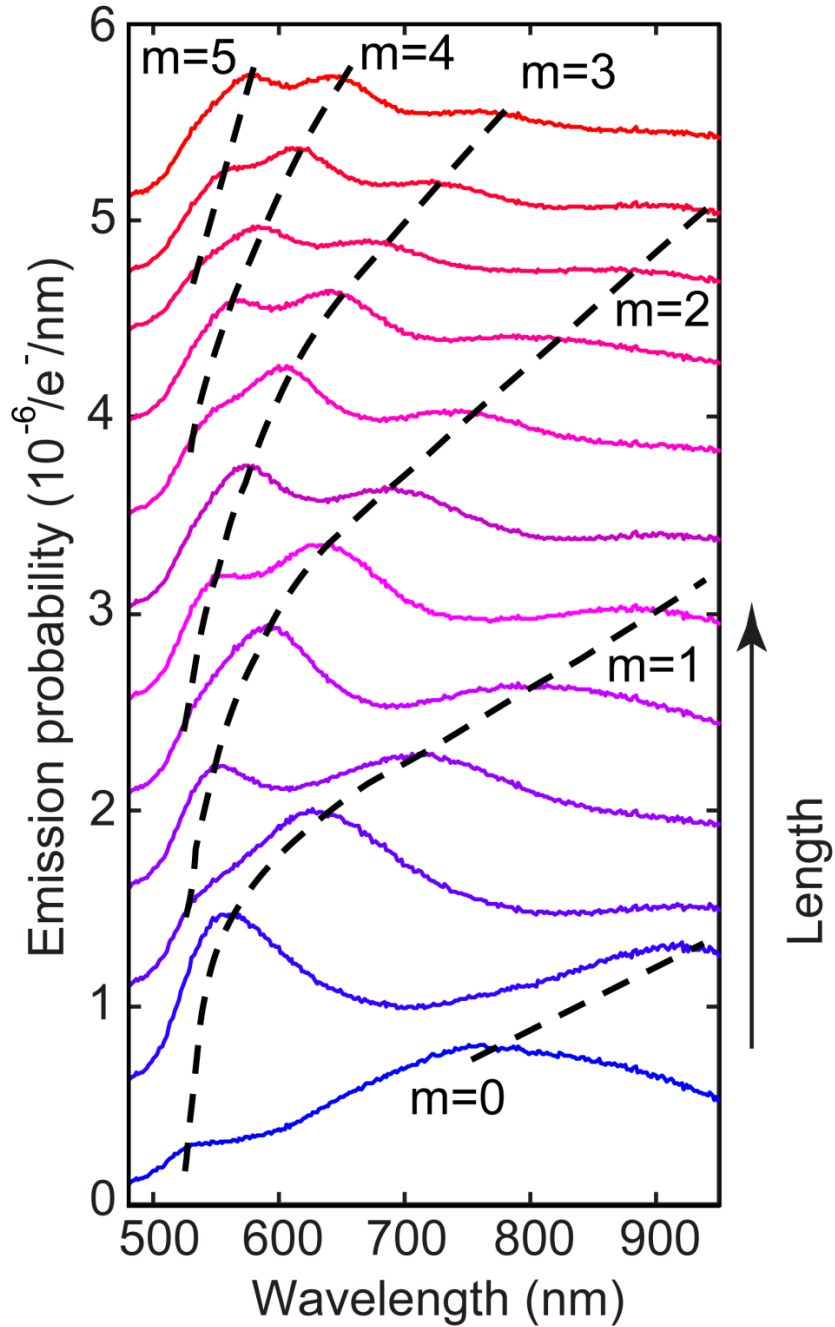

**Figure 3.** Integrated CL spectra for different ridge antennas with *L*=100-1200 (100 nm steps). The spectra have been vertically offset for clarity. The evolution of the resonance orders is indicated by the black dashed curves

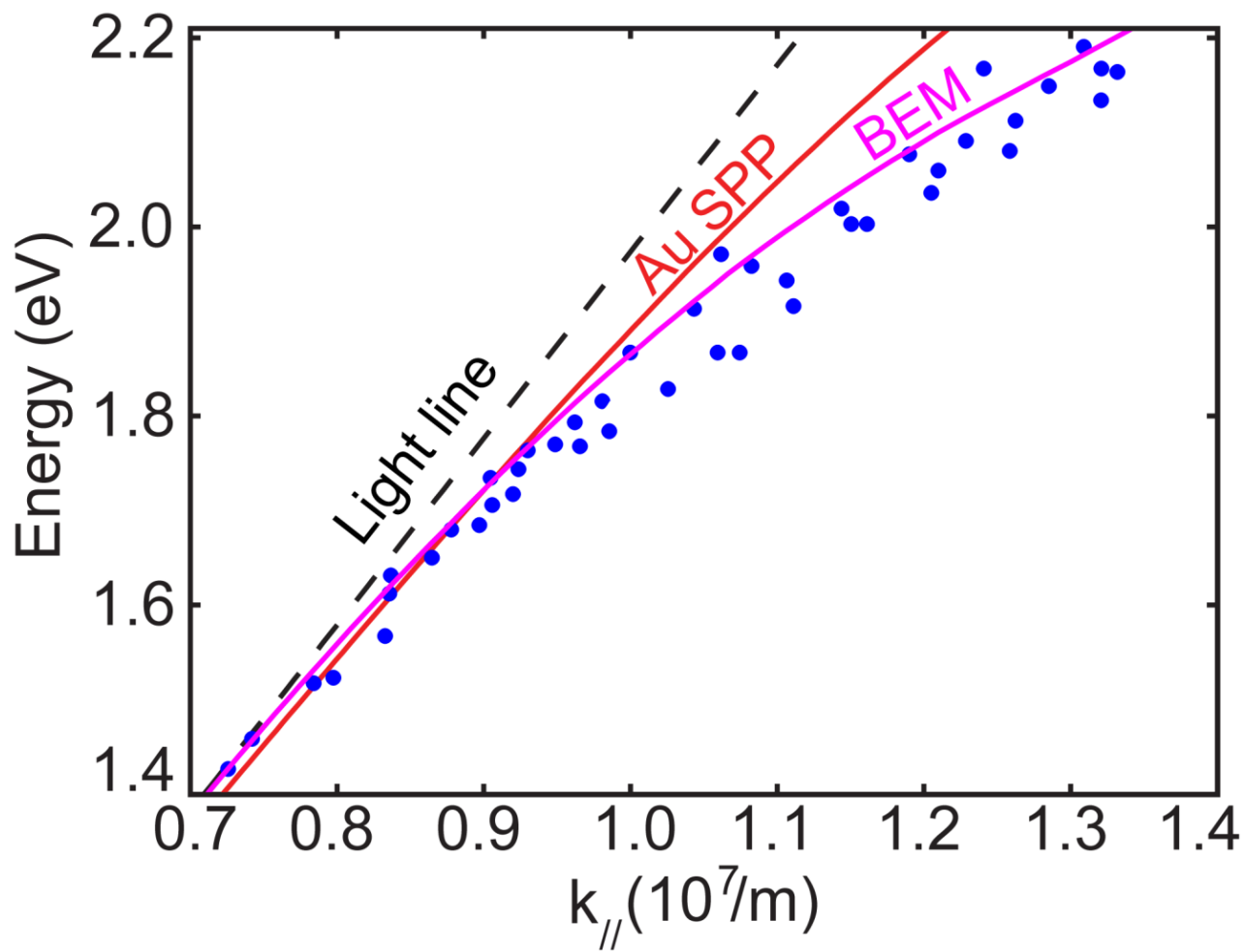

**Figure 4.** Experimental antenna dispersion (blue dots) derived from the standing wave patterns of ridges with *L*=700-2000 nm and a BEM dispersion calculation (magenta curve) showing plasmon wave vector *versus* energy. For reference, the light line in vacuum (black dashed curve) and the dispersion curve for a SPP on a planar single crystalline gold surface (red curve) are shown.

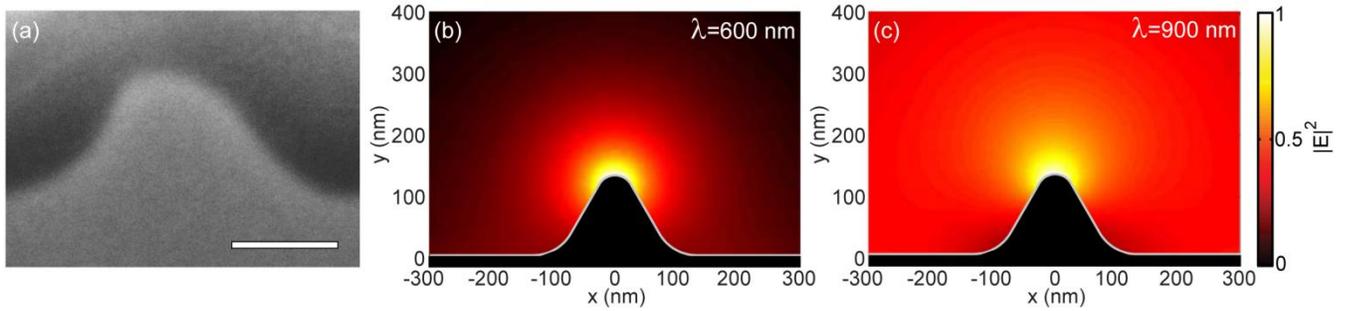

**Figure 5.** (a) SEM image taken of an antenna cross section made by FIB with a sample tilt of 52° (scale bar 100 nm). BEM calculation of the induced near field by a vertically oriented dipole positioned 10 nm above the ridge waveguide for (b) 600 nm at $k = 1.12k_0$ and (c) 900 nm at $k = 1.002k_0$. The ridge antenna geometry is indicated in black.

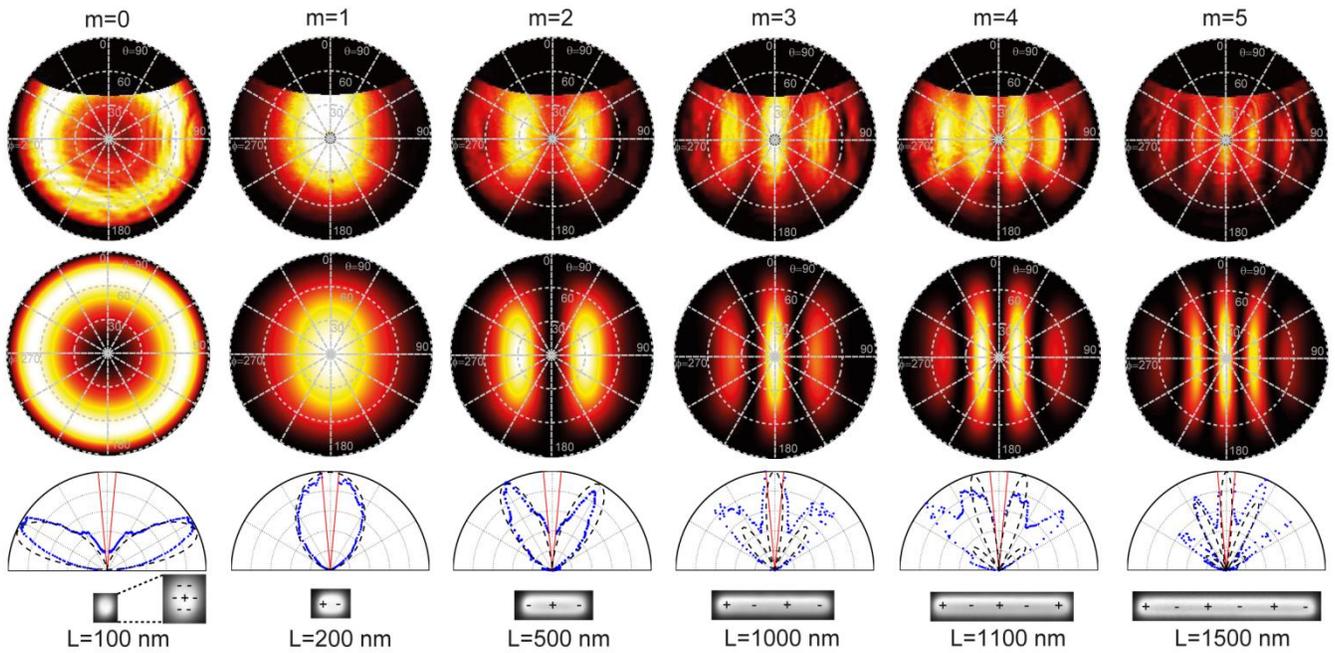

**Figure 6.** Measured angle-resolved emission patterns (top row), calculated emission patterns (second row). The antennas are oriented horizontally. The third row shows cross cuts through measurements (blue dots) and calculations (black dashed line) along $\varphi$=90° and $\varphi$=270°, showing emission patterns for $m$=0, 1, 2, 3, 4, and 5 resonances, collected from ridges with different lengths (see SEM images bottom row). In the SEM images we have also indicated snapshots of the charge distributions corresponding to each resonance (we have enlarged the SEM image for the 100 nm long antenna to improve the visibility). From left to right the measurements were collected at $\lambda$=750, 550, 600, 650, 600 and 650 nm. The angular range that is not collected by the mirror because of the hole is indicated by the red curves ($\theta$<5°).

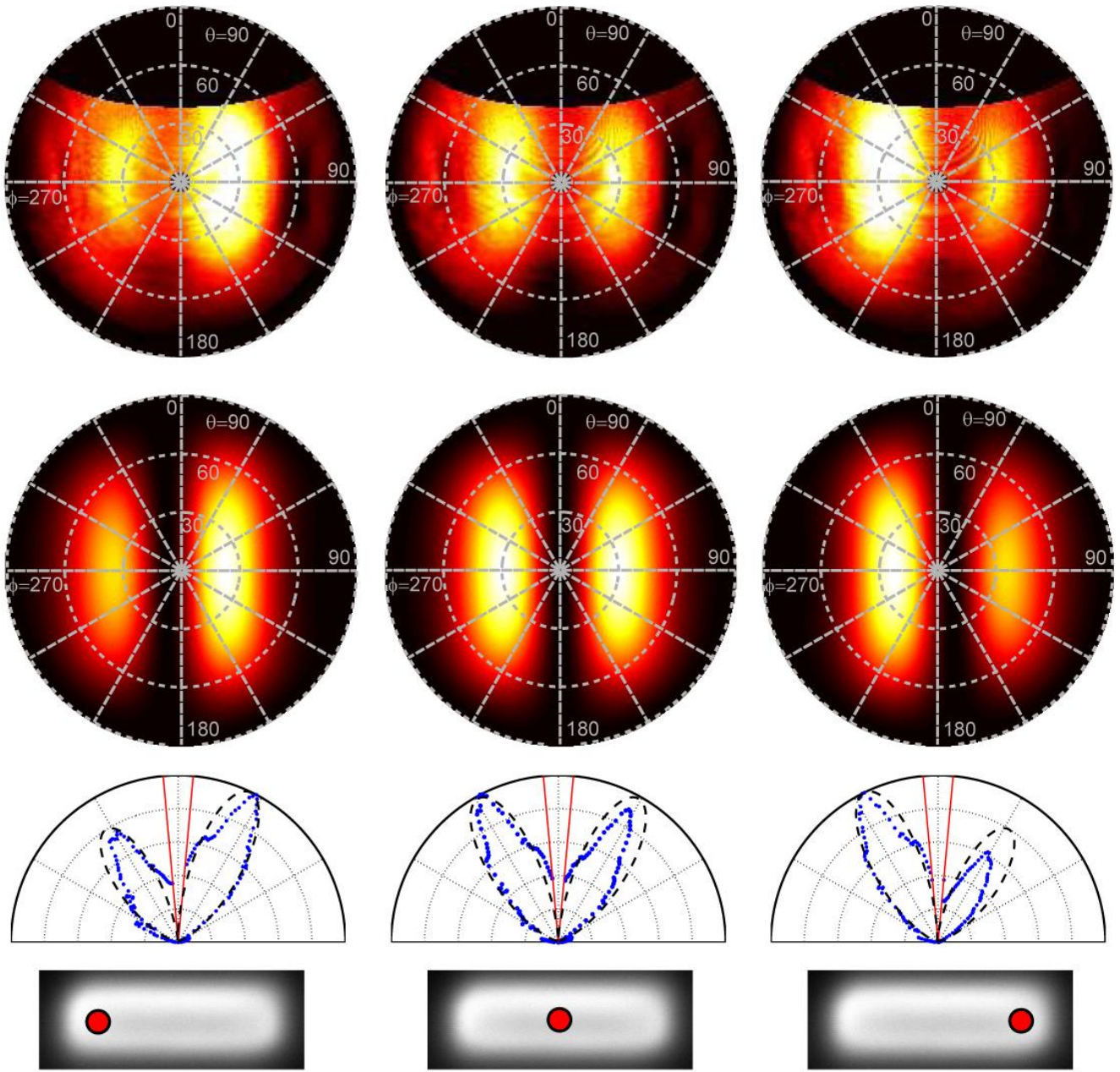

**Figure 7.** Measured angle-resolved emission patterns (top row), calculated emission patterns (second row) and cross cuts through both (third row) for excitation of the outer left, center and outer right position of the 500 nm long ridge (see bottom row), collected at $\lambda$= 600 nm.